\documentclass[%
 reprint,%
 amssymb, amsmath,%
 prc,cha,%
superscriptaddress
]{revtex4-2}
\usepackage[colorlinks=true,linkcolor=blue]{hyperref}%
\usepackage{physics}
\usepackage{graphicx}
\usepackage{lipsum}
\usepackage{mwe}
\usepackage{caption}

\begin{document}

\title{Clustering in nuclei at finite temperature}%

\author{Esra Y\"uksel}
\email{e.yuksel@surrey.ac.uk}
\affiliation{Department of Physics, University of Surrey, Guildford, Surrey GU2 7XH, United Kingdom}

\author{Florian Mercier}
\affiliation{IJCLab, Universit\'e Paris-Saclay, CNRS/IN2P3, 91405 Orsay Cedex, France}

\author{Jean-Paul Ebran}
\affiliation{CEA,DAM,DIF, F-91297 Arpajon, France}
\affiliation{Universit\'e Paris-Saclay, CEA, Laboratoire Mati\`ere en Conditions Extr\^emes, 91680, Bruy\`eres-le-Ch\^atel, France}

\author{Elias Khan}
\affiliation{IJCLab, Universit\'e Paris-Saclay, CNRS/IN2P3, 91405 Orsay Cedex, France}

\begin{abstract}
We investigate the localization and clustering features in $^{20}$Ne ($N=Z$) and neutron-rich $^{32}$Ne nuclei at zero and finite temperatures. The finite temperature Hartree-Bogoliubov theory is used with the relativistic density-dependent meson-nucleon coupling functional DD-ME2. It is shown that clustering features gradually weaken with increasing temperature and disappear when the shape phase transition occurs. Considering thermal fluctuations in the density profiles, the clustering features vanish at lower temperatures, compared to the case without thermal fluctuations. The effect of the pairing correlations on the nucleon localization and the formation of cluster structures are also studied at finite temperatures. Due to the inclusion of pairing in the calculations, cluster structures are preserved until the critical temperatures for the shape phase transition are reached. Above the critical temperature of the shape phase transition, the clustering features suddenly disappear, which differs from the results without pairing.
\end{abstract}

\date{\today}%

\maketitle

\section{Introduction}
Among the vast diversity of possible nontrivial arrangements for nucleons within nuclei, localization phenomena, and in particular nuclear clustering ---
the formation of nucleon bound structures within a nucleus --- is one of the most intriguing structural organization. 
The ignition of nuclear clustering in the low-density region of nuclear matter has been extensively studied~\cite{PhysRevC.95.045804, Schuck_2016,PhysRevLett.80.3177,SCHUCK2007285} and shown to play a decisive role in the nuclear equation of state (EOS), hence, in the modeling of astrophysical events~\cite{PhysRevC.77.055804,Fischer2014,Custodio2020}. How these features translate into finite nuclei, i.e., how the clusterization phenomenon manifests in nuclei and impacts their spectroscopic properties, represents an active area of research.

One of the first predictions of molecular states and $\alpha$ clustering of nuclei, dates back to 1930s \cite{RevModPhys.8.82, PhysRev.52.1107}. Although it is difficult to observe it directly from experiments, several studies have been performed to detect $\alpha$ clustering in nuclei \cite{VONOERTZEN200643,RevModPhys.90.035004,PhysRevLett.112.162501, PhysRevLett.113.012502}. From the theoretical point of view, a variety of microscopic models have also been used to detect cluster structures and analyze their formation in nuclei. For instance, antisymmetrized [fermion] molecular dynamics (AMD) [FMD] \cite{PhysRevC.52.628,PhysRevC.52.647,10.1143/PTPS.142.205, PhysRevLett.98.032501}, Brink and generator coordinate method (Brink-GCM) \cite{10.1143/PTP.57.1262,10.1143/PTP.62.1621},  Tohsaki-Horiuchi-Schuck-R\"opke (THSR) wave function \cite{PhysRevLett.87.192501}, and the nuclear energy density functional (NEDF) approach~\cite{PhysRevC.71.064308,PhysRevC.74.044311, PhysRevC.83.034312,Ebran2012, PhysRevC.87.044307, PhysRevC.89.031303, PhysRevC.90.054307} can be considered as the most important theoretical tools to study the clustering phenomenon. Among them, the NEDF approach stands as the most suitable way to study clustering features in finite nuclei throughout the nuclide chart~\cite{RevModPhys.75.121, 10.1088/2053-2563/aae0ed}. Within this framework, the origin of the localization and clustering in finite nuclei has been studied in Ref. \cite{Ebran2012} using the relativistic and non-relativistic NEDFs, and it has been shown that correlations responsible for deformation features play a major role in the formation of cluster states.  Using relativistic NEDF, the conditions of the formation of cluster structures in $N=Z$ finite nuclei are also discussed in Refs. \cite{PhysRevC.87.044307,PhysRevC.89.031303}.  Although cluster structures occur in light nuclei with $N \approx Z$, the studies also point out the formation of novel types of clusters in neutron-rich nuclei with the emergence of molecular bonds of $\alpha$ particles \cite{PhysRevC.52.628,PhysRevC.52.647,10.1143/PTPS.142.205,PhysRevC.71.064308,PhysRevC.83.034312, PhysRevC.90.054329}.

The clustering phenomenon is expected to appear in the low density, low temperature region of nuclear matter. Up to now, the temperature effects have been studied for the clustering of nucleons only in nuclear matter \cite{PhysRevLett.80.3177,PhysRevC.77.055804, PhysRevC.77.035806, PhysRevC.81.049902, PhysRevC.79.034604, PhysRevC.79.014002, PhysRevC.95.045804, Custodio2020}. Concomitantly, comprehensive studies have also been conducted using the relativistic and non-relativistic NEDFs to understand the properties and behavior of nuclei with increasing temperature \cite{GOODMAN198130, BONCHE1984278, PhysRevC.88.034308, Yuksel2014, PhysRevC.92.014302, PhysRevC.93.024321, PhysRevC.102.044328, YUKSEL2021122238,PhysRevC.104.064302}. However, there is no study on the temperature dependence of the localization and clustering features in finite nuclei, which is the aim of the present work.

The present article is organized as follows. In Sec. II, a brief description of the finite temperature relativistic Hartree-Bogoliubov (FT-RHB) method is given. In Sec. III, we first present the results for $^{20}$Ne, and then discuss the changes in the densities and clustering features with increasing temperature. Then, the effect of pairing correlations on the disappearance of clusters is discussed in both $^{20}$Ne and $^{32}$Ne. Finally, the conclusions are given in Sec. IV.

\section{Theoretical formalism}\label{sec:theoretical_formalism}
Within the finite temperature framework, the nucleus is considered as a grand canonical ensemble, and the equilibrium state is obtained
by applying the variational principle to the grand canonical potential of the system.
The latter reads 
\begin{equation}
\Omega = E -TS-\lambda N,
\label{grand}
\end{equation}
where $E$ is the total energy, $S$ is the entropy, and $N$ is the particle number. Also, $T$ and $\lambda$ represent temperature and chemical potential, respectively. 
The finite temperature equations are obtained by minimizing the grand canonical potential $\delta \Omega =0 $ (see Refs. \cite{GOODMAN198130,Egido_1993} for details). 

The FT-RHB equations have the same form as the zero-temperature case. The FT-RHB matrix is given by \cite{NIKSIC20141808}
\begin{equation}
\resizebox{0.9\hsize}{!}{$
\left(
\begin{array}{cc}
h_D-\lambda-m & \Delta \\
-\Delta^{*} & -h_D^{*}+\lambda+m
\end{array}
\right)
\left(\begin{array}{l}
U_k\\ V_k
\end{array}
\right)
= E_k \left(\begin{array}{l}
U_k \\ V_k
\end{array}
\right),
$}
\end{equation}
where $h_D$,  $\Delta$ and $m$ represent the single-nucleon Dirac Hamiltonian, pairing field, and nucleon mass, respectively. 
The chemical potential $\lambda$ is determined by the particle number subsidiary condition $\langle\hat{N}\rangle = \text{Tr}[\rho] = N$. The quasiparticle energies are denoted by $E_k$, and $U_k, V_k$ are the corresponding RHB wave functions. For the ground-state solution of an even-even nucleus, the Dirac Hamiltonian for the density-dependent meson-exchange model is given by
\begin{equation}
h_D= -i\boldsymbol{\alpha} \boldsymbol\nabla +V(\boldsymbol{r})+\beta(m+S(\boldsymbol{r})),
\end{equation}
where $S$ and $V$ represent the attractive scalar and repulsive vector potentials, respectively. More specifically, 
\begin{subequations}
\begin{align}
S(\boldsymbol{r}) &= g_{\sigma}\sigma, \\    
V(\boldsymbol{r}) &= g_{\omega}\omega+g_{\rho}\tau_{3}\rho+eA_{0}+\Sigma_{0}^{R},
\end{align}
\end{subequations}
where $g_{\sigma}$, $g_{\omega}$, and $g_{\rho}$ are the (density-dependent) vertex functions of the Lorentz-scalar and isoscalar bilinear forms of the nucleon operators, and $A_{0}$ is the time component of the electromagnetic field (see Ref. \cite{NIKSIC20141808} for the explicit forms of the equations). The rearrangement contribution due to the density dependence of the vertex functions is given by 
\begin{equation}
\Sigma_{0}^{R}=\frac{\partial g_{\sigma}}{\partial \rho_{v}}\rho_{s}\sigma+\frac{\partial g_{\omega}}{\partial \rho_{v}}\rho_{v}\omega+\frac{\partial g_{\rho}}{\partial \rho_{v}}\rho_{tv}\rho.
\end{equation}

At finite temperatures, the scalar, vector and isovector densities read
\begin{subequations}
\begin{align} 
\rho_{s} &=\sum_{E_{k}>0} V_{k}^{\dagger}\gamma^{0}\left(1-f_{k}\right) V_{k}+U_{k}^{T}\gamma^{0} f_{k} U_{k}^{*},\\
 \rho_{v} &=\sum_{E_{k}>0} V_{k}^{\dagger}\left(1-f_{k}\right) V_{k}+U_{k}^{T} f_{k} U_{k}^{*},\\
 \rho_{tv } &=\sum_{E_{k}>0} V_{k}^{\dagger} \tau_{3}\left(1-f_{k}\right) V_{k}+U_{k}^{T}\tau_{3} f_{k} U_{k}^{*}, 
 \end{align}
\end{subequations}
 where $f_k$ is the Fermi-Dirac function
\begin{equation}
f_{k}=\frac{1}{1+e^{\beta E_{k}}},
\end{equation}
and $\beta\equiv 1/k_{B}T$, with $k_B$ the Boltzmann constant. 

The pairing field is 
\begin{equation}
\Delta_{l l^{\prime}}=\frac{1}{2} \sum_{k k^{\prime}} V_{l l^{\prime} k k^{\prime}}^{p p} \kappa_{k k^{\prime}},
\end{equation}
where $V^{pp}$ is the matrix element of the particle-particle ($pp$) interaction~\cite{PhysRevC.69.054317,TIAN200944}. The pairing tensor $\kappa$ is defined as
\begin{equation}
\kappa = \sum \limits_{E_k > 0 } V_k^{*} (1-f_k) U_k^{T} +U_k f_k V_k^{\dagger}.
\end{equation}

For the particle-particle ($pp$) interaction $V^{pp}$,  we use the separable interaction of the form \cite{PhysRevC.69.054317,TIAN200944} 
\begin{equation}\label{eq-15}
V^{pp}\left(\boldsymbol{r}_{1}, \boldsymbol{r}_{2}, \boldsymbol{r}_{1}^{\prime}, \boldsymbol{r}_{2}^{\prime}\right)=-G \delta\left(\boldsymbol{R}-\boldsymbol{R}^{\prime}\right) P(\boldsymbol{r}) P\left(\boldsymbol{r}^{\prime}\right) \frac{1}{2}\left(1-P^{\sigma}\right),
\end{equation}
where $\boldsymbol{R} = \frac{1}{2}(\boldsymbol{r}_1 + \boldsymbol{r}_2)$ and $\boldsymbol{r} = \boldsymbol{r}_1 - \boldsymbol{r}_2$ are the center-of-mass and relative coordinates respectively, and $P(\boldsymbol{r})$ is defined as
\begin{equation}\label{eq-16}
P(\boldsymbol{r})=\frac{1}{\left(4 \pi a^{2}\right)^{3 / 2}} e^{-\frac{\boldsymbol{r}^{2}}{4 a^{2}}}.
\end{equation}
Unless otherwise stated, the parameters of the $pp$ interaction are taken as $G_{p(n)} = 728$ MeV.fm${}^3$ and $a=0.644$ fm for DD-ME2 \cite{PhysRevC.71.024312} interaction. The mean pairing gap $\Delta$ is defined as
\begin{equation}
\Delta = \frac{E_{pair}}{\Tr \kappa}=\frac{\sum_{l l^\prime} \Delta_{l l^\prime} \kappa_{l l^\prime}}{\sum_l \kappa_{ll}}.
\end{equation}

In this work, we solve the FT-RHB equations with an additional constraint in the axial mass quadrupole moment and minimize the grand canonical potential.
The density-dependent meson-exchange DD-ME2 functional \cite{PhysRevC.71.024312} is used in the calculations.
The quadrupole deformation $\beta_2$ spans the [$-$0.6,1.0] interval with steps of 0.05, to generate the Free energy surfaces. The equations of motion are expanded in an axial harmonic oscillator basis with 20 (48) harmonic shells, for fermionic (bosonic) quantities.
The Free energy of a nucleus is calculated with $F(\beta_{2},T)=E(\beta_{2},T)-TS(\beta_{2},T)$. Here, $E(\beta_{2},T)<0$ is the total energy of the system and $S(\beta_{2},T)$ is the entropy for a given deformation parameter $\beta_2$ at finite temperature $T$. The entropy is calculated as \cite{GOODMAN198130}
\begin{equation}
S= -k_{B} \sum_{k} \left[f_{k}\ln f_{k}+(1-f_{k})\ln (1-f_{k})\right].
\label{entropy}
\end{equation}
The internal excitation energy of the nucleus is given by $E^{*}(\beta_{2},T)=E(\beta_{2},T>0)-E(\beta_{2},T=0)$.

For a realistic modeling of nuclei at finite temperatures, both statistical (thermal) and quantum fluctuations need to be considered. Quantum fluctuations are known to be important for light nuclei and at low temperatures ($T<$1 MeV), and play a minor role for nuclei with a sharp minimum in their free energy surface \cite{Egido_1993,PhysRevLett.61.767}. Up to now, several studies have been performed to take into account quantum fluctuations \cite{PhysRevLett.61.767,PhysRevC.59.185,PUDDU1991409,PhysRevB.45.9882,ROSSIGNOLI1997242,ROSSIGNOLI1999719c}. In addition, in mesoscopic systems such as finite nuclei, statistical fluctuations around the minimum of the free energy are expected to yield a significant contribution at finite temperatures. In this work, we only consider the statistical fluctuations in the calculations.

In a first step, relevant finite-temperature properties, e.g., pairing gaps, quadrupole deformations, excitation energies, etc., can be computed from the lowest state in the free energy surface, characterized by a quadrupole deformation $\beta_{2}^\star$ at a given temperature.
To account for such effects, an observable $O$ is not computed from the state minimizing the free energy surface, i.e., as $O=O(\beta_2^\star,T)$, but after mixing expectation values $O(\beta_2,T)$ at various deformations $\beta_2$ under the form of the ensemble average $O=\overline{O}$~\cite{osti_4690472,PhysRevC.29.1887,PhysRevC.68.034327}, with 
\begin{equation}
\overline{O} = \frac{\int d\beta_{2} O(\beta_{2},T) \exp(-\Delta F(\beta_{2},T)/T)}{\int d\beta_{2} \exp(-\Delta F(\beta_{2},T)/T)}.
\label{ave}
\end{equation}

The exponential weight in Eq.~\eqref{ave} represents the probability $P(\beta_{2},T)$ to obtain a deformed configuration with a quadrupole parameter $\beta_2$ at temperature $T$~\cite{osti_4690472,PhysRevC.29.1887,PhysRevC.68.034327}, 
\begin{equation}
P(\beta_{2},T) \propto \exp(-\Delta F(\beta_{2},T)/T),
\label{P}
\end{equation}
where $\Delta F(\beta_{2},T)$ is the Free energy relative to the lowest state in the free energy surface, for a given deformation and temperature:
\begin{equation}
\Delta F(\beta_{2},T)=F(\beta_{2},T)-F(\beta_{2}^\star,T).
\label{free}
\end{equation}

According to Eq. \eqref{P}, there are no thermal fluctuations at zero temperature, and the impact of the fluctuations increases with temperature.

\section{Results}
\subsection{$^{20}$Ne nucleus}
\label{1}

\begin{figure}
\centering
\includegraphics[width=\linewidth]{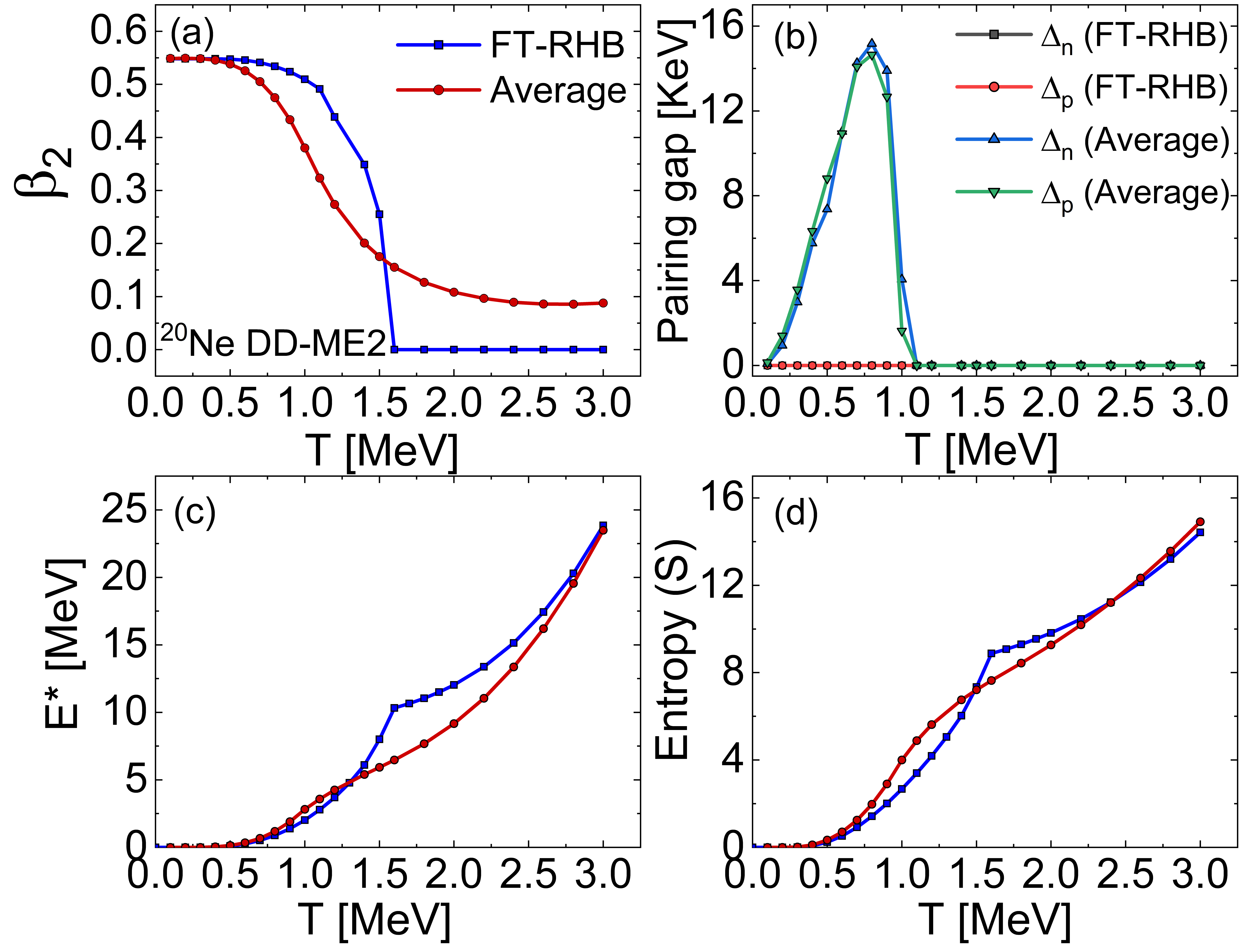} 
\caption{The variation of the most probable values of the deformation (a), pairing gap (b), excitation energy (c), and entropy (d) as a function of temperature. The calculations are performed using the FT-RHB and DD-ME2 functional. The average values of the observables are calculated using the constrained FT-RHB results at finite temperatures [see Eq. (\ref{ave})]}. \label{fig:1}
\end{figure}

\begin{figure*} 
 \centering
\includegraphics[width=\linewidth]{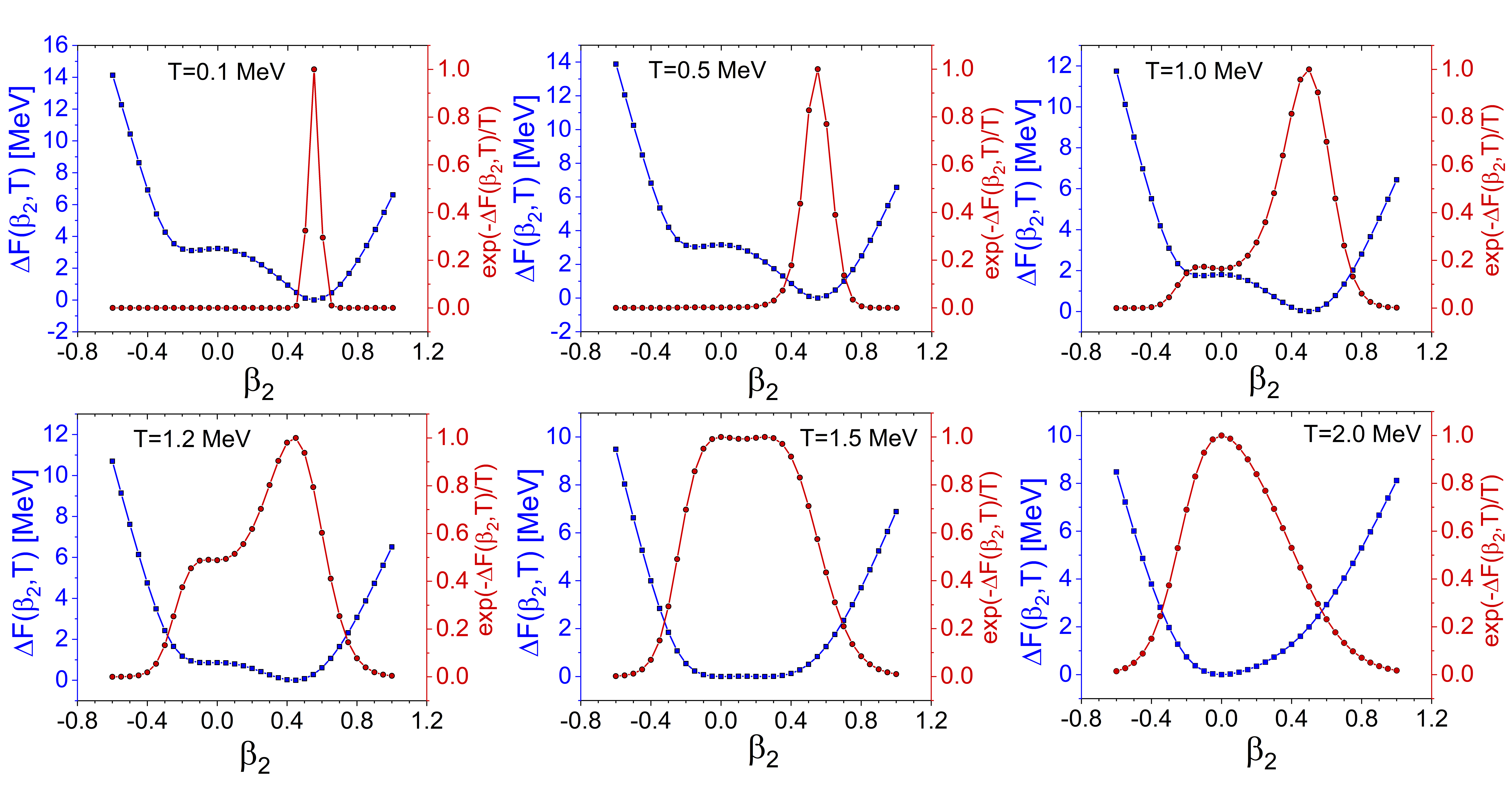} 
\caption{Relative free energies $\Delta F(\beta_{2},T)$ (blue line with a square symbol) as a function of deformation at finite temperatures. The calculations are performed using the constrained FT-RHB with the DD-ME2 functional for $^{20}$Ne, and the relative free energies are calculated using Eq. \eqref{free} (see the text for the details). The probability factors exp$(-\Delta F(\beta_{2}, T)/T)$ are also shown on the right side of the y-axis using red lines with a circle symbol.}\label{fig:2}
\end{figure*}

$^{20}$Ne is a well-known nucleus that displays a strong intrinsic quadrupole deformation, accompanied by a pronounced localization in intrinsic densities \cite{PhysRevC.89.031303,VONOERTZEN200643, Ebran2012, PhysRevC.87.044307}. To illustrate the effect of the temperature on the properties of $^{20}$Ne, we perform the FT-RHB calculations using the DD-ME2 functional. In Fig. \ref{fig:1}, the FT-RHB results are presented for the deformation, pairing gap, excitation energy, and entropy. 
At zero temperature, the ground-state of $^{20}$Ne is predicted to have a highly deformed prolate shape with $\beta_{2} =0.55$ and the total binding energy is obtained as 157.64 MeV. The results are in agreement with the experimental data, where the $\beta_{2}$ deformation and the total binding energy are obtained as 0.720(20) and 160.64 MeV \cite{Wang_2021}, respectively. Considering only the FT-RHB minimum of the Free energy at finite temperatures, it is seen that the deformation decreases with increasing temperature and a shape phase transition from prolate to spherical is obtained just above $T=1.5$ MeV (see Fig. \ref{fig:1}(a)). The decrease in the deformation is mainly caused by the depopulation of the deformation-driven states and the disappearance of shell effects at high temperatures \cite{GOODMAN198130, PhysRevC.93.024321}.

To analyze the effect of thermal shape fluctuations at finite temperature, we perform constrained FT-RHB calculations for fixed $\beta_{2}$ values from $-$0.6 to $+$1.0 with steps of 0.05. Then, the deformation-dependent Free energy surfaces are obtained and thermal averages of the observables are calculated according to Eq. \eqref{ave} and displayed in Fig. \ref{fig:1}. As expected, taking into account the thermal shape fluctuations in the calculations, the sharp shape phase transition is removed and the decrease in the $\beta_{2}$ value becomes smoother with increasing temperature. Although its value remains low ($\beta_{2}\simeq$ 0.1), deformation does not disappear even at $T=3.0$ MeV. The authors in Ref. \cite{PhysRevC.59.185} also performed calculations for $^{20}$Ne by taking into account the quantum fluctuations, albeit using a different Hamiltonian. In contrast to our results, it has been shown that the deformation remains almost constant at all temperatures with the inclusion of the quantum correlations.

To better understand the effect of thermal fluctuations, in Fig.~\ref{fig:2}, we display the Free energy surfaces and the probability factors as a function of the deformation at finite temperatures. The free energy surfaces are obtained using Eq. \eqref{free} and the probability factors are calculated using Eq. \eqref{P}. According to Eq. \eqref{P}, the probability factor depends on the free energy and temperature with $\exp(-\Delta F(\beta_{2},T)/T)$. At $T=0.1$ MeV, the free energy surface has a well-formed minimum around $\beta_{2}=0.55$. Hence, the probability factor is found to be 1 for this state and zero for others. As temperature increases from 0.1 to 1 MeV, the contribution of other states to the thermal average of an observable increases because of the changes in the free energy surfaces. At higher temperatures, the free energy surface becomes flat, and the contribution of other states to the thermal average of observables increases, as seen from the changes in the probability factors. Because of the increasing contribution of other states to the thermal average of the observables with increasing temperature, we obtain a smooth decrease in the deformation.

We also analyze the changes in the pairing gap, excitation energy, and entropy with increasing temperature. The proton and neutron pairing gap values are predicted to be zero using the FT-RHB, whereas we obtain a very small pairing gap value since we take the average of pairing over the ensemble of possible quadrupole shapes. This gap starts to develop at low temperatures and vanishes before $T=1.0$ MeV [see Fig.\ref{fig:1} (b))].

The internal excitation energy increases with increasing temperature, and a visible kink around the shape phase transition temperature within FT-RHB is predicted. Using the thermal averages method, the internal excitation energy also increases with temperature, but the kink disappears and the change in the energy becomes smoother. Similar results are also obtained for the entropy (see Fig.\ref{fig:1}(c)-(d)).

\begin{figure}
    \centering
\includegraphics[width=\linewidth]{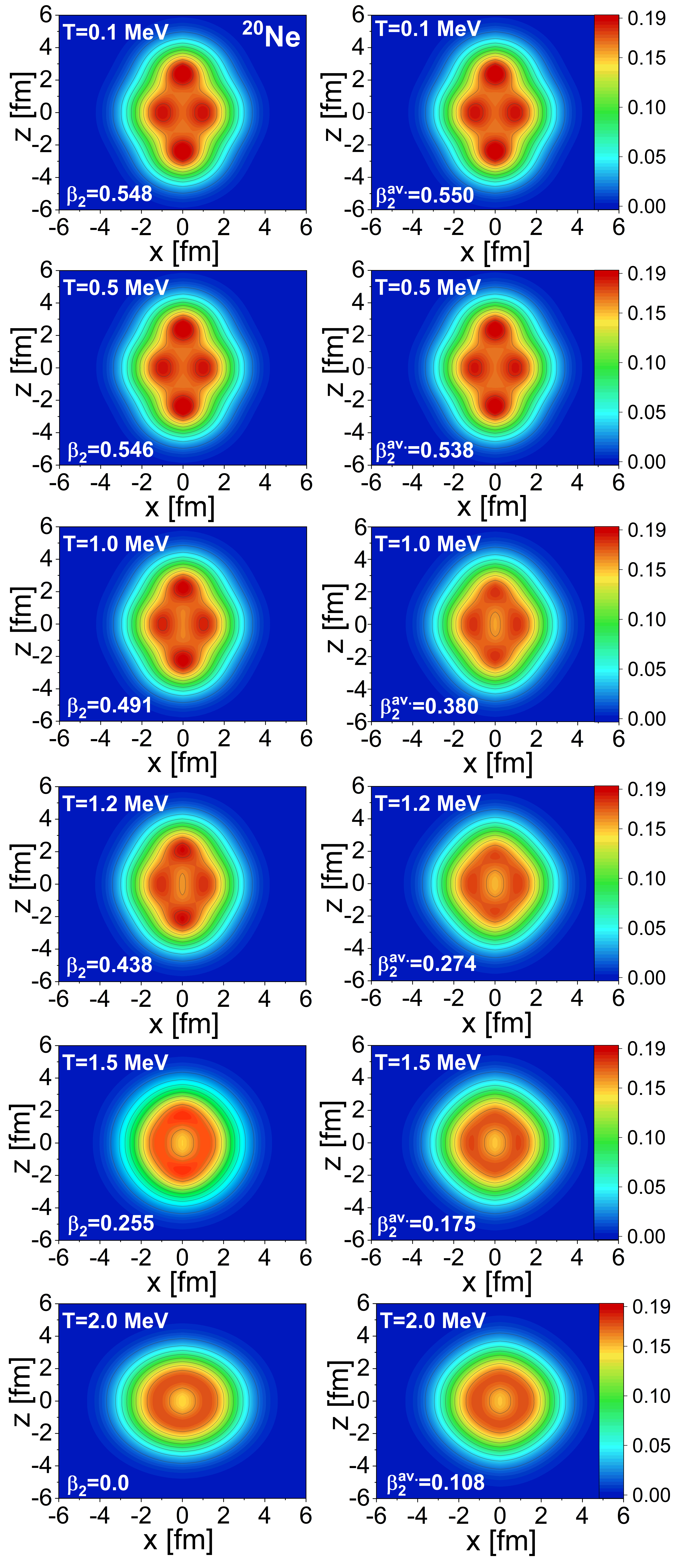} %
\caption{Left panels: The total intrinsic density ($\rho_{n}+\rho_{p}$) (in units of fm$^{-3}$) for $^{20}$Ne using DD-ME2 functional at finite temperatures. Right panels: The thermal average of the total density using Eq. \eqref{ave}. The quadrupole deformation values, $\beta_2$ and $\beta_{2}^{av.}$, are displayed in the bottom corners of the panels. }\label{fig:3}
\end{figure}

 We now study the changes in the localization properties and behavior of $\alpha$ clusters in $^{20}$Ne with increasing temperature. In the left panels of Fig. \ref{fig:3}, we display the total intrinsic nucleon densities in the $x-z$ plane at finite temperatures associated with the FT-RHB minimum of the free energy. It is seen that the DD-ME2 functional predicts a localized density profile and $\alpha$ clusters are visible at $T=0.1$ MeV, as expected \cite{Ebran2012}. At low temperatures, the density profile remains almost the same, with presence of $\alpha$ clusters. Increasing temperature further and reaching $T=1$ MeV, the density profile starts to change and localization becomes weaker. After $T=1$ MeV, $\alpha$ clusters start to disappear and vanish completely before the shape phase transition. After $T=1.5$ MeV, the nucleons display the characteristics of a Fermi liquid. 

At finite temperatures, the changes in the density profile and clustering features of finite nuclei are mainly related to the changes in the deformation properties of nuclei. By increasing temperature, the nucleons scatter to the high energy states, which in turn leads to an increase (decrease) in the occupation probabilities of the states above (below) the Fermi level \cite{PhysRevC.88.034308,PhysRevC.93.024321, YUKSEL2021122238}. Therefore, the Fermi surface smears, shell effects weaken, and deformation decreases with increasing temperatures \cite{PhysRevLett.85.26, PhysRevC.96.054308,PhysRevC.93.024321}. Also, the wave functions are more spread, and the density profile is broadened through the surface region, which in turn, destroys the localization of the wave functions and clustering features of finite nuclei.

We also performed the constrained FT-RHB calculations at finite temperatures and calculated thermal averages of total intrinsic nucleon densities over the ensemble of quadrupole shapes at finite temperatures using Eq. \eqref{ave} (right panels of Fig. \ref{fig:3}). In the FT-RHB approach, the diagonal elements of the density matrix, or the density operator, give the probability density of finding a particle at a specific point \cite{PhysRevC.53.2809}, and one can calculate density as a function of deformation to calculate the thermal average of the density at finite temperatures.
By doing so, the thermal average for the density is associated with finding the average probability densities at each point in the $x-z$ plane over the ensemble of quadrupole shapes at finite temperatures. Using the thermal average densities to pinpoint $\alpha$ clusters, the calculations predict the disappearance of localized patterns earlier and at lower temperatures, compared to the unconstrained FT-RHB results. Although the changes in the density profiles are the same at low temperatures, cluster states disappear completely at $T=1$ MeV using the thermal average densities in the analysis. This feature is due to the thermal mixing of various shapes (see Fig. \ref{fig:3}), leading to a smoother density profile.

\begin{figure}
  \centering
\includegraphics[width=\linewidth]{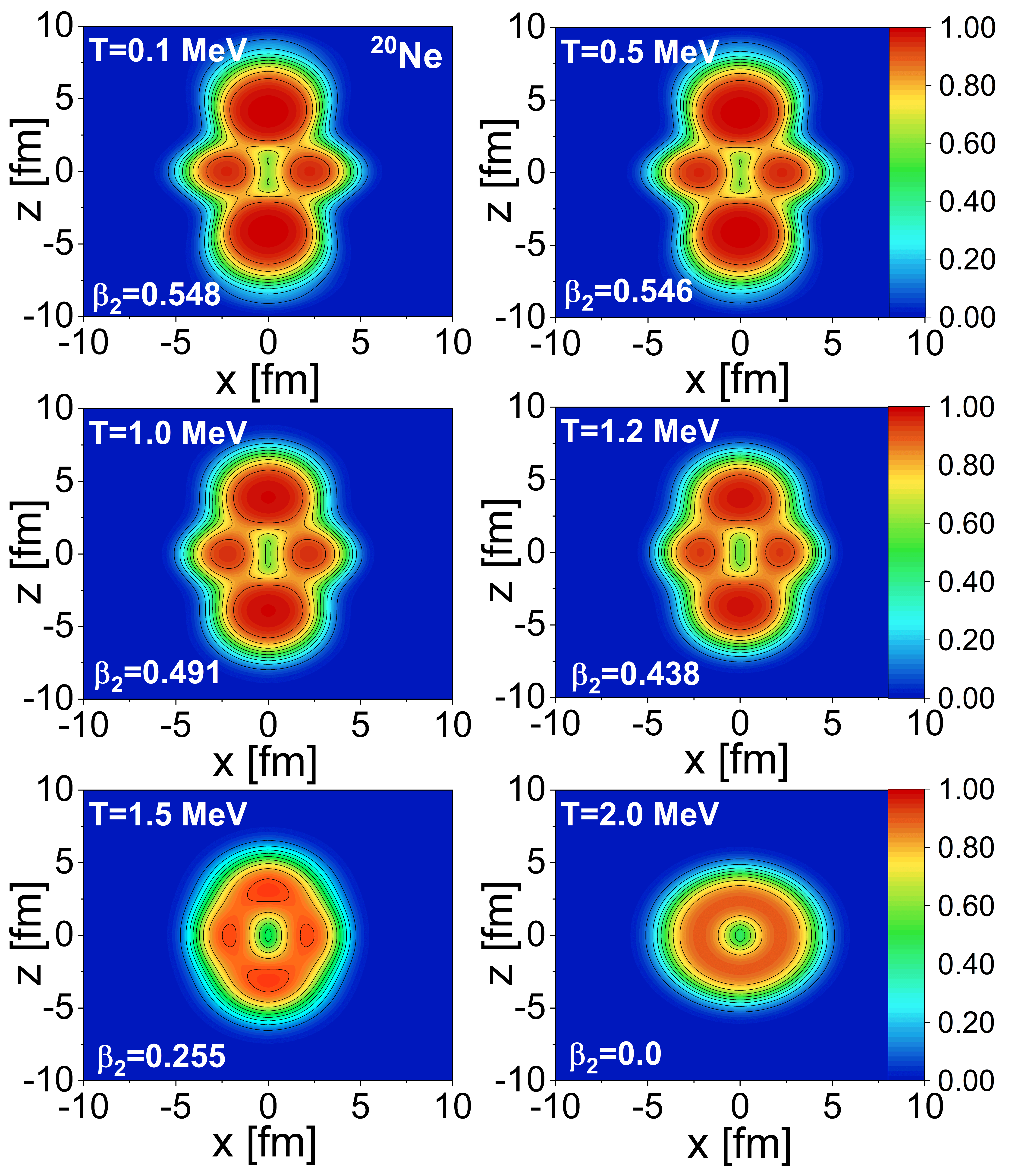} 
  \caption{The nuclear localization function for $^{20}$Ne using DD-ME2 functional at finite temperatures. The quadrupole  deformation $\beta_2$ values are displayed in the bottom corners of the panels.}\label{fig:4}
\end{figure}

Besides using the nucleon densities to detect the cluster structures in nuclei, the nucleon localization function also provides an alternative way to study them. Using the nucleon densities to study the clustering only contains information from mass density, whereas the nucleon localization function takes contributions from the kinetic energy, density, and density gradients, and provides a better measure for the localization of nuclei. Therefore, to illustrate the nucleon localization and the cluster structures in light nuclei, we also use the nucleon localization function (NLF) (see Refs. \cite{PhysRevC.83.034312, PhysRevC.94.064323}). The nucleon localization is expressed by
\begin{equation}
 {\mathcal C}_{q\sigma}(\mathbf{r})=
  \left[1+\left(\frac{\tau_{q\sigma}\rho_{q\sigma}-\frac{1}{4}[\pmb{\nabla}\rho_{q\sigma}]^2}
       {\rho_{q\sigma}\tau_{q\sigma}^\mathrm{TF}} \right)^2 \right]^{-1},
\label{loc}
\end{equation}

where $\tau_{q\sigma}$, $\rho_{q\sigma}$, and $\nabla \rho_{q\sigma}$ are kinetic energy density, particle density, and density gradient, respectively. The Thomas–Fermi kinetic energy density in the denominator is given by $\tau^{TF}=\frac{3}{5}(6\pi^{2})^{2/3}\rho_{q\sigma}^{5/3}$.
The $\mathcal C_{q\sigma}(\textbf{r})$ function provides a dimensionless and normalized measure of nucleon localization. The likelihood of finding two nucleons with the same spin and isospin at the same spatial position is low when $C$ is close to one. Because of the weak Coulomb interaction for light $N = Z$ nuclei, proton and neutron localization functions are similar, whereas they need to be studied separately for neutron-rich nuclei, as we discuss below. In this work, the $\alpha$ localization function is calculated with $\mathcal C_{\sigma}=\sqrt{\mathcal{C}_{p\sigma}\mathcal{C}_{n\sigma}}$.

The localization function results for $^{20}$Ne are displayed in Fig.~\ref{fig:4} at finite temperatures. As we mentioned above, when C is close to one, $\alpha$-like clusters (a pure overlap of four nucleon wave functions) are expected because of the spin and isospin degeneracy. Compared to the density profiles, the localization function typically displays a larger spatial extension because of the kinetic term included in the calculations. At low temperatures, the localization function predicts highly localized regions at the outer ends and the $\alpha$ clusters are clearly visible for $^{20}$Ne. The localization feature is also obtained around the center, which is associated with $^{12}$C. Therefore, this pattern is interpreted as $\alpha-$$^{12}$$\text{C}-\alpha$ quasimolecular configuration \cite{PhysRevC.83.034312,PhysRevC.89.031303}. Similar to the findings above, the cluster structures almost do not change up to $T=1$ MeV. Increasing temperature further, the clustering effects start to weaken and completely disappear before the shape phase transition of the nucleus, in agreement with the findings using the total intrinsic densities above.

\subsection{Pairing, deformation, and clustering in nuclei}
\label{2}

\begin{figure}
 \centering
\includegraphics[width=\linewidth]{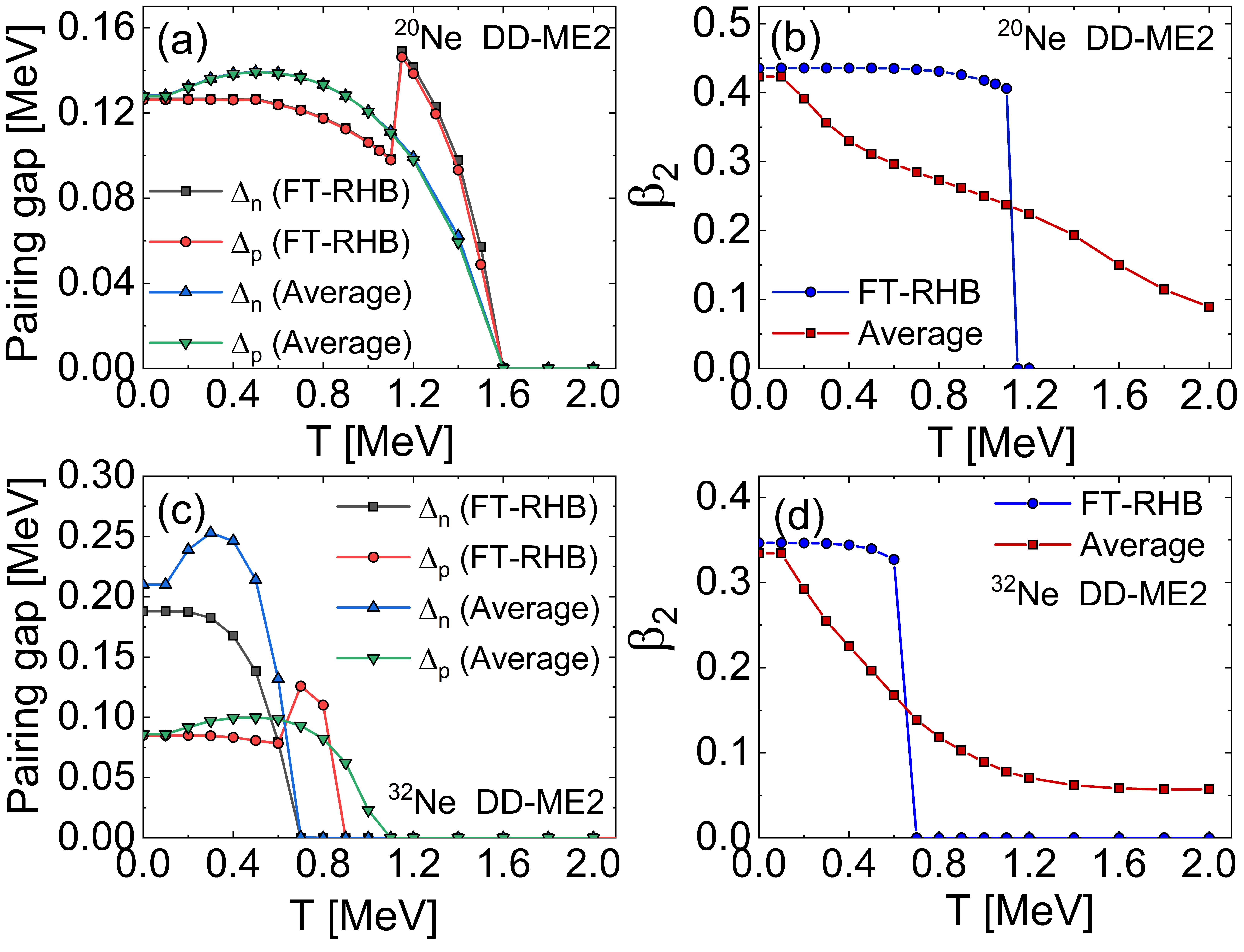} 
  \caption{The change of the pairing gap values for $^{20}$Ne (a) and $^{32}$Ne (c) as a function of  temperature. The change of deformation for $^{20}$Ne (b) and $^{32}$Ne (d) as a function of temperature. The averages of the observables are calculated using Eq. \eqref{ave}}.
  \label{fig:5}
\end{figure}

\begin{figure}
    \centering
\includegraphics[width=\linewidth]{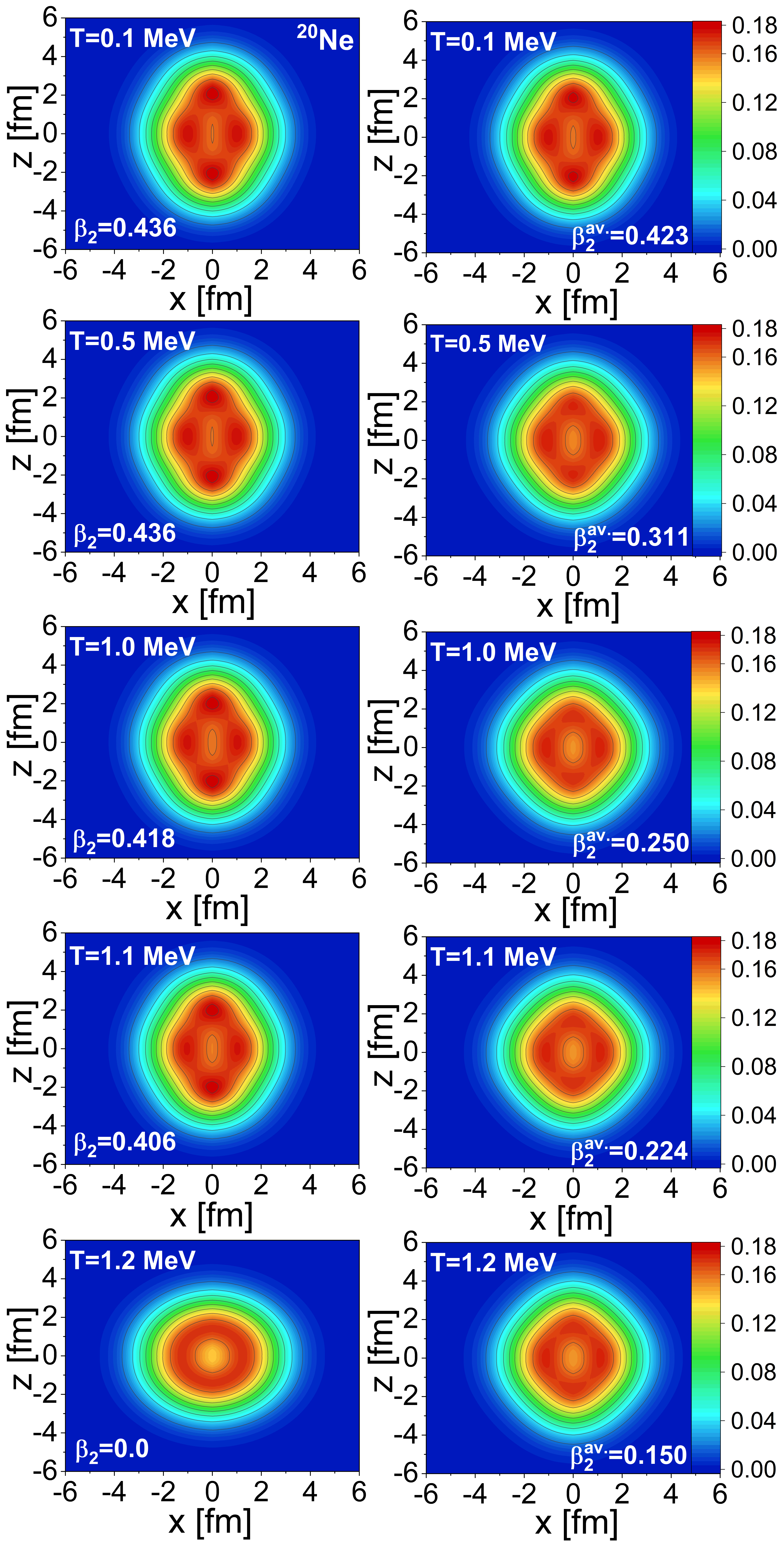} %
\caption{The same as in Fig. \ref{fig:3}, but using larger pairing strength, $G_{p(n)} = 910$ MeV.fm${}^3$, in the FT-RHB calculations for $^{20}$Ne. }\label{fig:6}
\end{figure}

\begin{figure}
  \centering
\includegraphics[width=\linewidth]{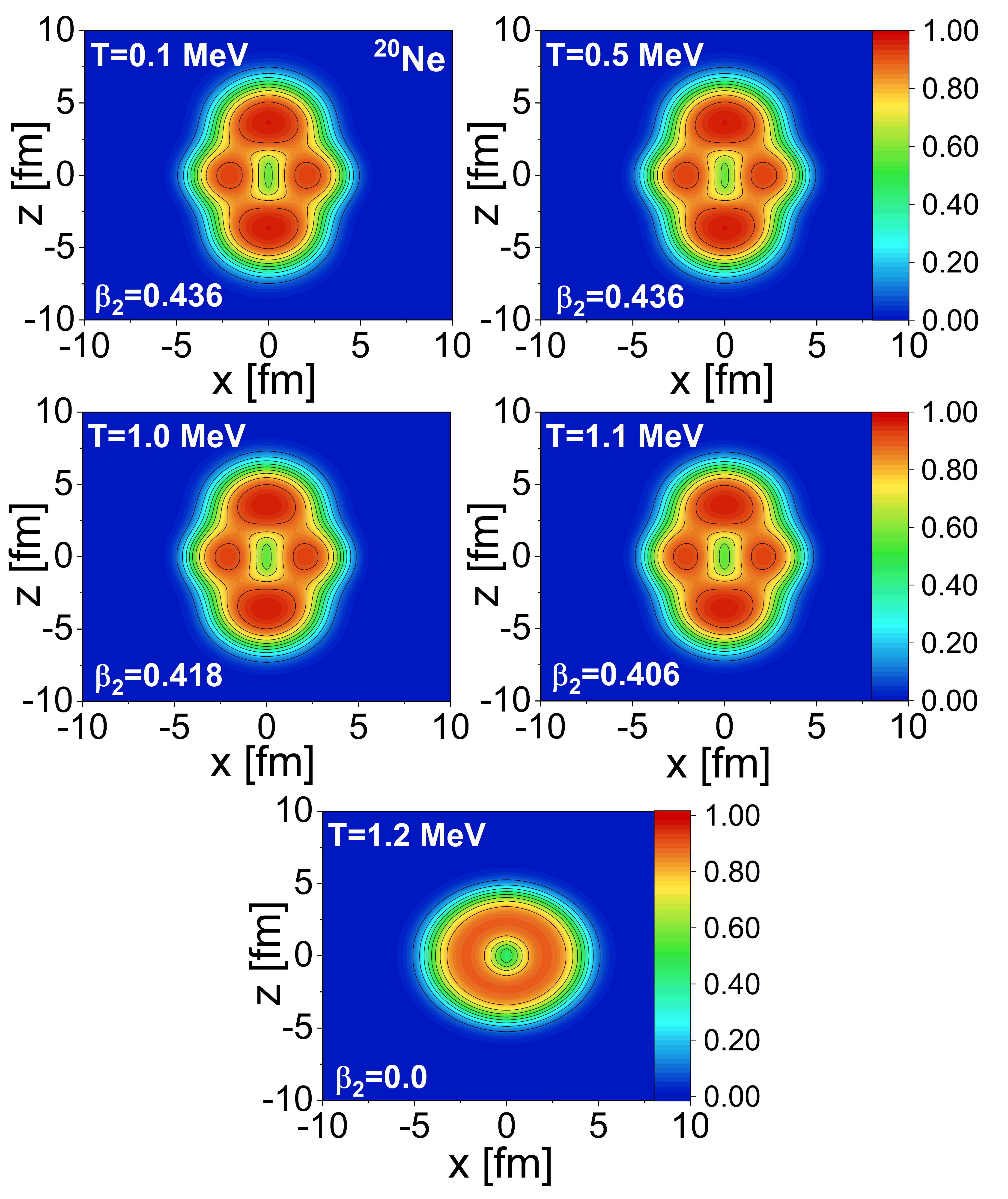} 
  \caption{The same as in Fig. \ref{fig:4}, but using larger pairing strength, $G_{p(n)} = 910$ MeV fm${}^3$ in the FT-RHB calculations.}\label{fig:7}
\end{figure}

\begin{figure*}
	\centering
	\includegraphics[width=\linewidth]{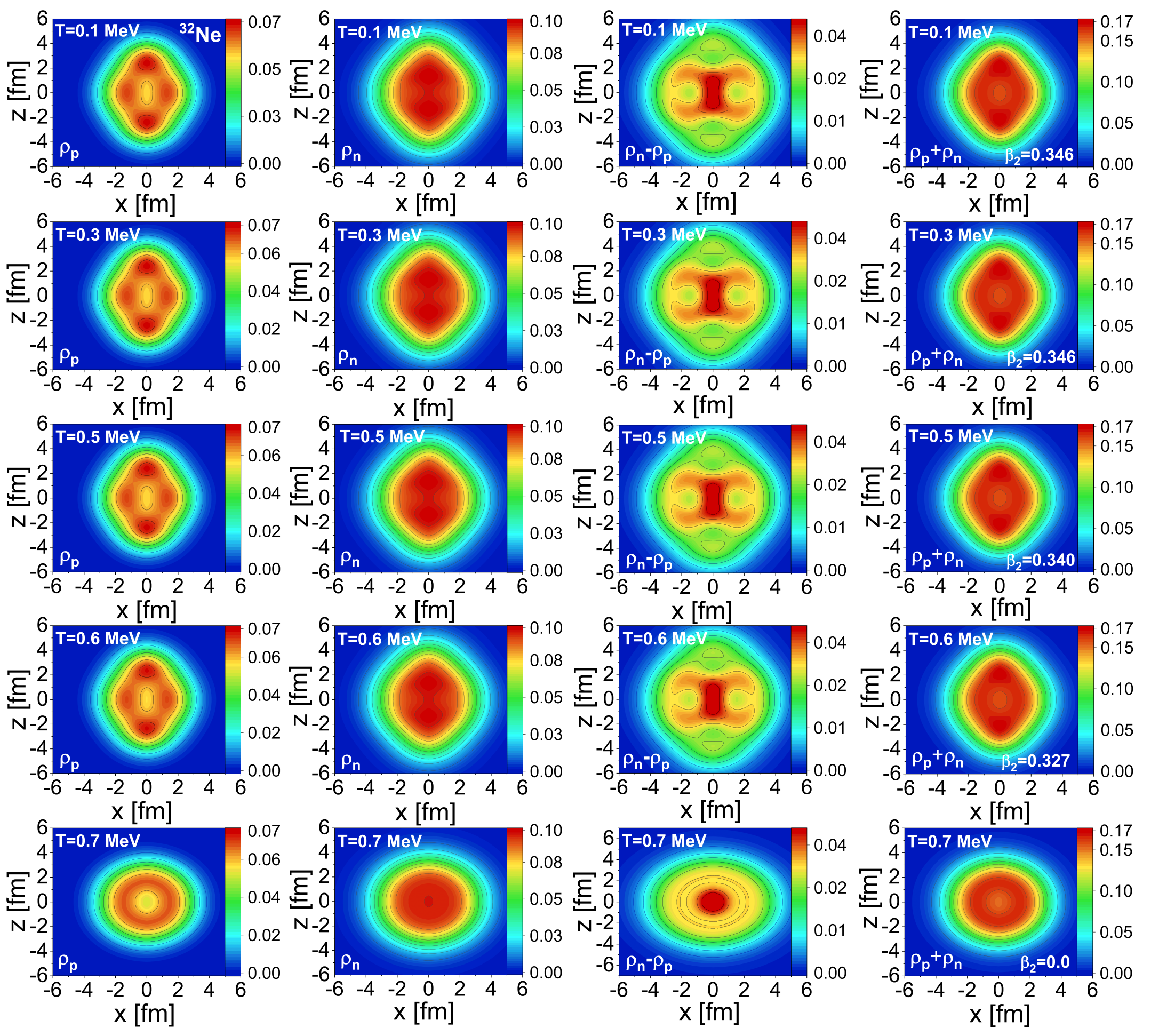} %
	\caption{The proton $\rho_{p}$, neutron $\rho_{n}$, isovector $\rho_{n}-\rho_{p}$, and isoscalar $\rho_{n}+\rho_{p}$ densities  (in units of fm$^{-3}$) for $^{32}$Ne. The calculations are performed using the FT-RHB and DD-ME2 functional at finite temperatures.  In the bottom corners of the rightmost panels, the quadrupole deformation $\beta_2$ values are displayed at a given temperature.}\label{fig:9}
\end{figure*}

\begin{figure*}
	\centering
	\includegraphics[width=\linewidth]{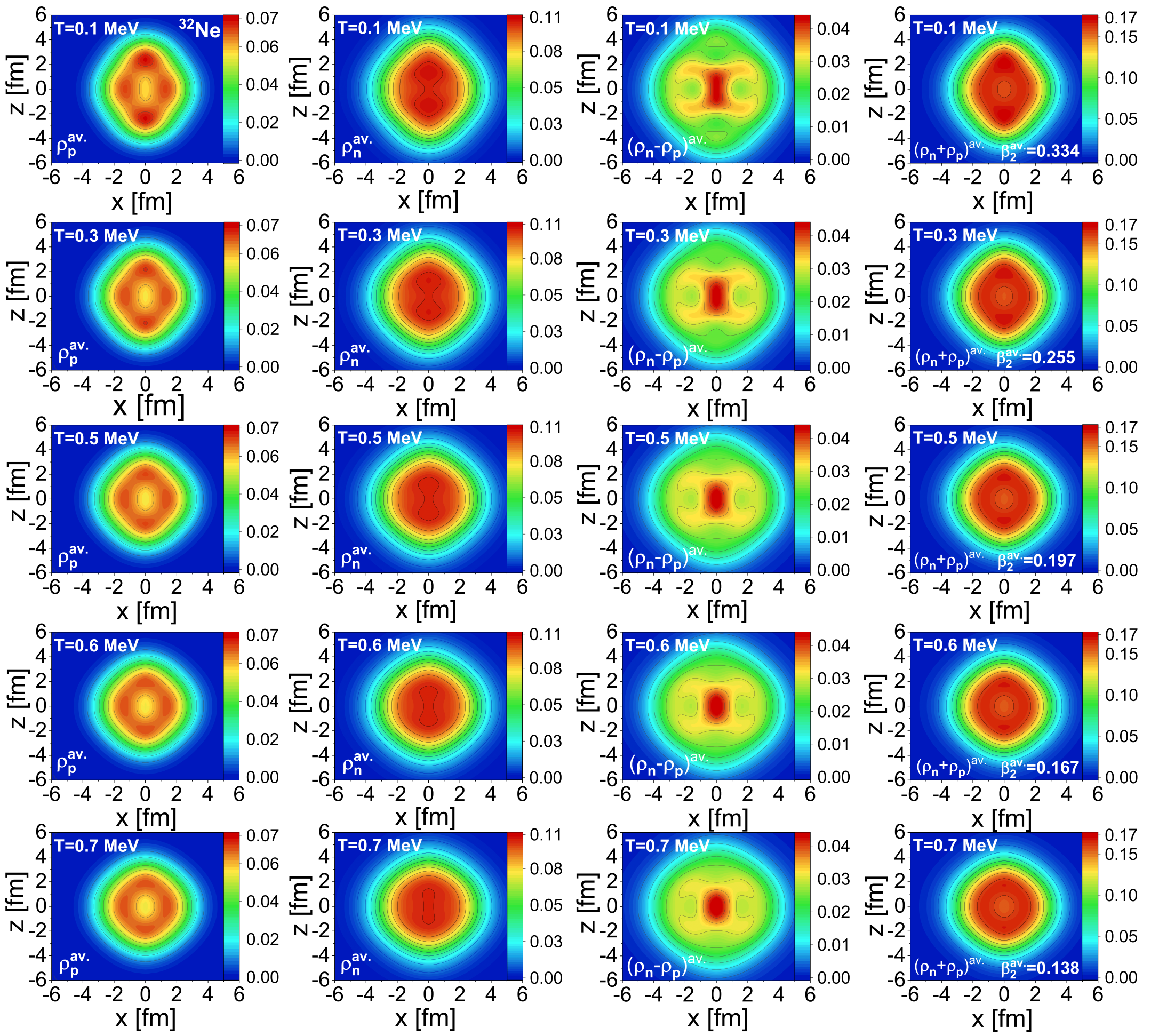} %
	\caption{The thermal average of the proton $\rho_{p}$, neutron $\rho_{n}$, isovector $\rho_{n}-\rho_{p}$, and isoscalar $\rho_{n}+\rho_{p}$ densities (in units of fm$^{-3}$) for $^{32}$Ne. The calculations are performed using the FT-RHB and DD-ME2 functional at finite temperatures.  In the bottom corners of the rightmost panels, the average of the quadrupole deformation  $\beta_{2}^{av.}$ value is displayed at a given temperature.}\label{fig:10}
\end{figure*}

In this section, we study the role of the pairing correlations in the disappearance of the localization properties and cluster states of finite nuclei with increasing temperature. For this purpose, we first perform calculations for $^{20}$Ne with increased pairing strength, $G_{p(n)} = 910$ MeV fm${}^3$ ($25\%$ increase), to induce a finite amount of pairing in this nucleus while maintaining the cluster structures. It should be noted that further increasing the pairing strength makes the deformation properties decrease considerably, and the localization features disappear. The $\alpha$ clustering can also be present in neutron-rich nuclei (see Refs. \cite{RevModPhys.90.035004,PhysRevC.90.054329}). Therefore, we also choose another neutron-rich nucleus, $^{32}$Ne, which has both pairing and clustering features, and use the standard value $G_{p(n)} = 728$ MeV fm${}^3$ as the pairing strength. Although the clustering features shall be weak in $^{32}$Ne compared to $^{20}$Ne, it allows studying the interplay between the pairing, deformation and clustering.

Let us first discuss the properties of $^{20}$Ne with increased pairing strength. In panels (a) and (b) of Fig. \ref{fig:5}, we display the pairing gap values and deformation for $^{20}$Ne as a function of the temperature. At zero temperature, the FT-RHB minimum for the free energy yields a highly deformed configuration at $\beta=0.45$ for $^{20}$Ne, despite the artificially increased pairing strength. As expected, the FT-RHB calculations predict a decrease in the pairing gap values and deformation with temperature. Although the pairing effects are small for $^{20}$Ne, they persist even after $T>$1 MeV, and superfluid-to-normal phase transition occurs after the shape phase transition [see Fig. \ref{fig:5} (a)' (b)]. A slight decrease and then a sudden increase in the pairing gap values at $T=1.1$ MeV is predicted. Increasing temperature further, pairing effects continue to decrease and completely vanish at $T=1.6$ MeV. The sudden increase in the pairing gap values is related to the change in the deformation of the nucleus after $T=1.1$ MeV, due to the shape phase transition. It is known that the single (quasi)-particle states are not sensitive to the changes in the temperature for the considered temperature range, whereas the changes in the deformation considerably affect these levels \cite{PhysRevLett.85.26,PhysRevC.93.024321, PhysRevC.96.054308}. The deformation removes the degeneracy of the single-particle levels and also plays an important role in the formation of clusters in nuclei \cite{Ebran2012}. With increasing temperature, the shape phase transition of the nucleus occurs after $T$ $>$1.1 MeV, and single-particle levels around the Fermi levels become degenerate because of the sudden disappearance of the deformation. As a result, the distribution of the nucleons in the single-particle levels changes, and we obtain an increase in the pairing gap values [see Fig. \ref{fig:5} (a)]. The reentrance of pairing at finite temperatures has also been discussed in Ref. \cite{PhysRevC.86.065801}. Looking at the thermal averages of the observables over the ensemble of quadrupole shapes, both the pairing gap and deformation decrease with increasing temperature. While the pairing gaps vanish at around $T=1.6$ MeV, there is still a small amount of deformation at high temperatures.

In Fig. \ref{fig:6}, we display the total intrinsic nucleon densities (left panels) and the thermal averages of the densities over the ensemble of quadrupole states (right panels) for $^{20}$Ne at finite temperatures. At $T=0.1$ MeV, clustering features can be seen, albeit they are faint compared to the findings in left panels of Fig. \ref{fig:3}. Starting with the total nucleon densities, the localization of nucleons and cluster structures remains almost the same up to the critical temperatures for the shape phase transitions, namely, up to $T=1.1$ MeV. However, a sudden change in the clustering features is obtained after $T=1.1$ MeV, i.e., the signals of cluster states disappear. This behavior differs from the findings in Fig. \ref{fig:3}, in which the cluster structures weaken gradually with increasing temperature rather than disappearing suddenly. Using the thermal averages of the densities, clustering features are not distinctively visible at $T=0.1$ MeV compared to the total intrinsic nucleon densities. They vanish with a slight increase in temperature. The nuclear localization function results are also displayed in Fig. \ref{fig:7}, providing a clear picture of the $\alpha$ clusters. Similar to the findings using the total nucleon densities, the cluster structures remain almost the same until the critical temperatures for the shape phase transitions, and then disappear suddenly. This behavior differs from the results in Fig. \ref{fig:4}, in which we do not have pairing effects and localization features disappear gradually with increasing temperature. Compared to the findings in Sec. \ref{1}, inclusion of the pairing effects at finite temperature keeps the deformation almost constant until the critical temperatures and delays the disappearance of the cluster structures up to the critical temperatures for the shape phase transition of nuclei.

Now, we investigate the ground-state and localization properties of neutron-rich $^{32}$Ne using the DD-ME2 functional. In panels (c) and (d) of Fig. \ref{fig:5}, the changes in the the pairing gap and the quadrupole deformation values are presented as a function of temperature. At zero temperature, the FT-RHB minimum of the free energy yields a configuration with $\beta=0.35$. Also, the total binding energy is obtained as 214.84 MeV and is in good agreement with the experimental data (213.47 MeV) \cite{Wang_2021}. While pairing decreases slightly, deformation stays almost constant up to $T=0.6$ MeV. Just after $T=0.6$ MeV, the shape phase transition occurs, and the nucleus becomes spherical. Concomitantly, the proton pairing is slightly increased. It should be noted that in $^{32}$Ne, the order of appearance of critical temperatures for pairing and deformation is reversed, compared to the case of $^{20}$Ne. The critical temperatures for the pairing and shape phase transitions are also lower, compared to $^{20}$Ne, due to the excess of neutrons. Taking into account thermal fluctuation with Eq. \eqref{ave}, the deformation smoothly decreases with increasing temperature. We discuss below the consequences of these changes on the localization properties of nuclei.

For neutron-rich nuclei, the cluster states can usually be described by molecular bonding of $\alpha$-particles by the excess neutrons \cite{10.1143/PTPS.E68.464, 10.1143/PTPS.142.205, 10.1143/PTPS.192.1, PhysRevC.90.054329}. In this picture, the valence neutrons can form $\pi$-like or $\sigma$-like molecular bonding with the $\alpha$ cluster core. Since the proton and neutron numbers are not equal in $^{32}$Ne, the wave functions are also different, and the interpretation of the cluster states requires the analysis of the nucleon densities separately. To this aim, the proton $\rho_{p}$, neutron $\rho_{n}$, isovector $\rho_{n}-\rho_{p}$, and isoscalar $\rho_{n}+\rho_{p}$ densities are displayed separately in Fig. \ref{fig:9} at equilibrium deformation. As seen from Fig. \ref{fig:9}, the proton and neutron density distributions are quite different at $T=0.1$ MeV: while the proton density distributions display more localized structures in the outer edges of the nucleus, the neutron density displays the characteristics of the Fermi liquid due to the excess neutrons. The localization of nucleons can also be seen by looking at the isovector $\rho_{n}-\rho_{p}$ density distributions in the third column of Fig. \ref{fig:9}. The low density regions at the edges of the nucleus show the cluster structures for this nucleus. Looking at the total density distribution in the last column of Fig.  \ref{fig:9}, the observation of the cluster states is strongly hindered by the excess neutrons. With increasing temperature, the deformation decreases slightly and the density profiles almost do not change up to $T=0.6$ MeV. At higher temperatures, the nucleus goes through a sharp phase transition from deformed to spherical shape, and the localization features disappear in the density profiles.

In Fig. \ref{fig:10}, we also display the thermal averages of the densities over the ensemble of the different quadrupole shapes at finite temperatures. The calculations are performed using the constrained FT-RHB and DD-ME2 functional. Similar to the findings above, the localization features can be seen in the proton and isovector density distributions, whereas we cannot observe any pronounced localization in the neutron and total density distributions. By taking the thermal averages of the densities, the localization features disappear earlier compared to the findings above. Already at $T=0.3$ MeV, the localization features disappear completely.

\begin{figure}
    \centering
\includegraphics[width=\linewidth]{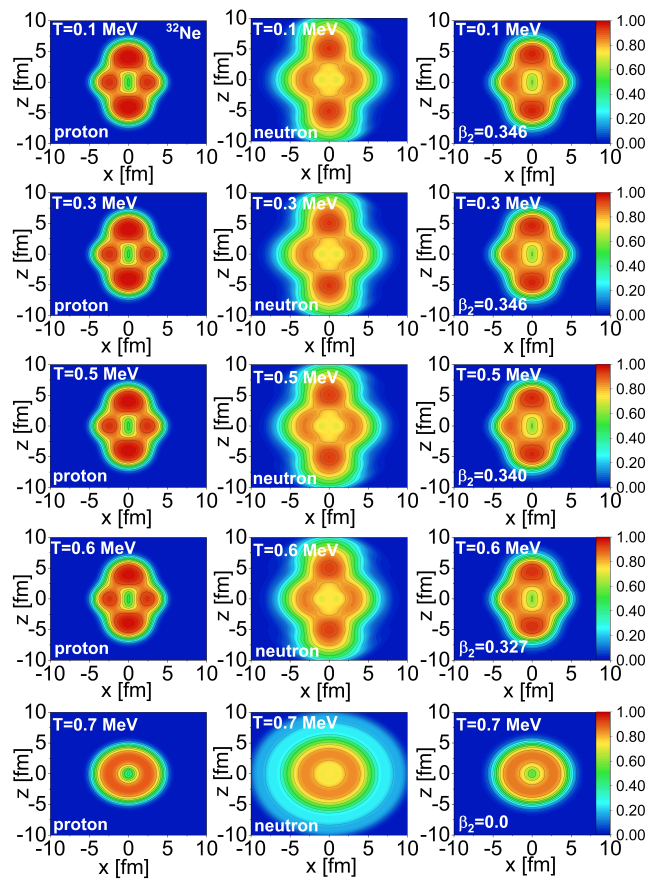} %
\caption{The proton, neutron and $\alpha$ localization $(C_{\sigma}=\sqrt{\mathcal{C}_{p\sigma}\mathcal{C}_{n\sigma}})$ functions for $^{32}$Ne. The calculations are performed using the FT-RHB and DD-ME2 functional at finite temperatures. In the bottom corners of the rightmost panels, the quadrupole deformation $\beta_2$ values are displayed for each temperature.}
\label{fig:8}
\end{figure}

The nuclear localization functions can provide a complementary picture for the clustering in neutron-rich nuclei. Therefore, in Fig. \ref{fig:8} we display the nuclear localization function as proton $(\mathcal{C}_{p\sigma})$, neutron $(\mathcal{C}_{n\sigma})$ and $\alpha$ localization functions $(C_{\sigma}=\sqrt{\mathcal{C}_{p\sigma}\mathcal{C}_{n\sigma}})$. At $T=0.1$ MeV, proton localization functions display highly localized regions at the end of the outer edges and around the center of  $^{32}$Ne, and the results look very similar to the $^{20}$Ne. As expected, the neutron localization function differs from the proton due to the neutron excess: it displays localization features only at the outer edges of $^{32}$Ne, which are also faint compared to the proton localization. Looking at $\alpha$ localization function, clustering structures are only apparent at the outer edges of the nucleus and are weaker compared to the proton localization due to the excess neutrons. Although we do not obtain any localization features using the total density (see the rightmost panels of Fig. \ref{fig:9}), the localization function results indicate clustering features, albeit weak. Similar to the findings in Fig. \ref{fig:9}, the localization functions almost do not change up to the critical temperature for the shape phase transitions. As soon as the nucleus becomes spherical, the cluster states disappear.

\section{Conclusions}
In this work, the FT-RHB calculations are performed to study the localization and clustering phenomenon in $^{20}$Ne ($N=Z$) and neutron-rich $^{32}$Ne nuclei at zero and finite temperatures. The relativistic density-dependent meson-nucleon coupling functional DD-ME2 is used in the calculations since it predicts much more localized density distributions and allows us to study the changes in nuclei with increasing temperature.

The constrained FT-RHB calculations predict a highly deformed ground-state and no pairing for $^{20}$Ne at zero temperature. The $^{20}$Ne nucleus displays a localized density profile and clustering features can be seen easily, both using the total intrinsic densities and nuclear localization functions. Performing unconstrained FT-RHB calculations, the localization and clustering features fade away gradually with increasing temperature and disappear completely with the shape phase transition of nuclei. Deformation removes the degeneracy of the single-particle states and plays a significant role in the formation of the cluster structures. Temperature weakens the deformation and spreads the density through the surface region, which in turn destroy the localization features in nuclei.

We also performed quadrupole constrained FT-RHB calculations at finite temperatures, and the thermal averages of some observables (deformation, pairing, entropy, excitation energy) and densities are also calculated over the ensemble of the different quadrupole shapes. Taking thermal averages of the observables, it is seen that the changes become smoother and a finite amount of deformation remains even at high temperatures. Analyzing thermal averages of the densities, we found that the clustering features disappear at lower temperature, compared to the unconstrained FT-RHB calculations.

We also study the effect of the pairing correlations on the disappearance of the clustering features at finite temperatures. By inducing a small amount of pairing for $^{20}$Ne, deformation and clustering features stay almost the same until the critical temperatures for the shape phase transitions. At higher temperatures and above $T=1.1$ MeV, the nucleus becomes spherical and clustering features disappear. An increase in the pairing, i.e., the reentrance of the pairing correlations, is predicted with the shape phase transition of the nucleus.
The neutron-rich $^{32}$Ne nucleus also displays localization and clustering features at zero temperature. We analyzed the localization of nucleons by decomposing the density profiles as well as localization function. Although the excess neutrons blur the observation of the cluster states, intrinsic proton densities and localization functions display rather pronounced cluster structures. Similar to the findings for $^{20}$Ne with pairing, the localization features keep the same shape until the critical temperatures for the phase transitions. For both $^{20}$Ne and $^{32}$Ne, the clustering features disappear at lower energy in the case of the thermal averages of the densities.

We should also mention that triaxiality can play an important role in the description of the ground-state properties and clustering features of some nuclei. Although the results may not change qualitatively, we also plan to extend our study in the future by including additional degrees of freedom in the calculations.

\bibliographystyle{apsrev4-2}
\bibliography{bibl}

\end{document}